# Optical parametric amplification of spatiotemporal optical vortices

Houan Teng[1,4], Chenhao Wan[2,4*], Qiwen Zhan[1,3*]

[1] Zhejiang Key Laboratory of 3D Micro/Nano Fabrication and Characterization, Department of Electronic and Information Engineering, School of Engineering, Westlake University, Hangzhou, China
[2] School of Optical and Electronic Information and Wuhan National Laboratory for Optoelectronics, Huazhong University of Science and Technology, Wuhan, Hubei 430074, China
[3] Westlake-Kromoson Joint Laboratory of Light-Field Intelligence, School of Engineering, Westlake University, Hangzhou 310030, China
[4] These authors contributed equally: Houan Teng, Chenhao Wan
*wanchenhao@hust.edu.cn
*zhanqiwen@westlake.edu.cn

## Abstract

**Spatiotemporal optical vortices (STOVs) are structured light featuring a phase singularity in the spatiotemporal domain, exhibiting unique topology that couples spatial and temporal dimensions. To date, amplification of STOVs remains unexplored, and their wavelength tunability is limited by available light sources. In this work, we present the theoretical analysis and experimental demonstration of optical parametric amplification (OPA) of STOVs. The pulse energy of the STOV signal is enhanced by a factor of 60, while its spatiotemporal structure is robustly preserved. The generated idler inherits the phase-conjugated STOV topology with the opposite topological charge, governed by simultaneous conservation of energy, linear momentum, and spatiotemporal orbital angular momentum. The intrinsic wavelength-conversion capability of OPA further provides a route towards spectrally tunable STOV generation, spanning from visible to near-infrared wavelength regimes. This work paves the way toward further development of multi-stage amplification architectures, facilitating applications of wavelength-tunable spatiotemporally structured light at ultrahigh intensities.**

## 1. Introduction

Structured light, underpinning modern nanooptics and photonics, can have intricate structures correlated in both space and time[1–4]. The past decade has witnessed rapid progress in spatiotemporally structured light, with research expanding from spatiotemporal optical vortices (STOVs) carrying transverse orbital angular momentum (OAM)[5,6], group-velocity-tunable spatiotemporal wave packets (STWPs)[7,8] and optical toroidal vortices[9,10], to optical skyrmions and other topological quasiparticles[11,12]. In particular, programmable spatio-spectral pulse shaping has enabled deterministic generation of STOVs with controllable topological structures[5,13], while advanced spatiotemporal interferometric techniques allow direct reconstruction of their complete amplitude and phase distributions[5,14,15]. These developments have established a powerful toolbox for generating, manipulating and characterizing increasingly complex spatiotemporal topology. The abundant and distinctive physical properties inherent to such fields enable a broad spectrum of promising applications covering extreme light–matter interactions[16,17], light-sheet microscopy[18], laser machining[19], optical information encoding and photonic computing[20–22].

Power scalability and wavelength tunability, however, remain two critical challenges towards high-field applications of spatiotemporally structured light. Considering the damage threshold and wavelength-dependent limitations of spatio-spectral shaping tools such as spatial light modulators (SLMs), it is preferable to generate intricate spatiotemporal structures within low-energy pulses and subsequently amplify and spectrally translate them. For STOVs, this is not simply a conventional gain problem: their topology is encoded in a coupled spatiotemporal complex field, and their transverse OAM depends on both phase winding and spatiotemporal field geometry. Optical parametric amplification (OPA), combining coherent phase-preserving amplification with high gain and intrinsic wavelength conversion[23,24], therefore provides a particularly attractive route to meet the challenges. It is worth exploring whether a pre-structured STOV can undergo strong stimulated parametric amplification while preserving its complete spatiotemporal topology, and how this topology is coherently redistributed between the amplified signal and the newly generated idler.

In this work, we present, to our knowledge for the first time, theoretical analysis and experimental demonstration of OPA of STOVs. As illustrated in Fig. 1, a seed beam at 780 nm selected from supercontinuum white light is structured as STOVs and amplified by a pump beam at 515 nm. The pulse energy of STOV signals is enhanced by approximately 60-fold while their spatiotemporal phase singularities and topological charges are robustly preserved. The generated idler inherits the phase-conjugated topology of the signal and carries the opposite topological charge. The simultaneous conservation of energy, linear momentum, and spatiotemporal orbital angular momentum govern the signal amplification and spatiotemporal topology transfer. The wavelength-conversion capability of OPA provides the wavelength tunability from visible to near-infrared wavelength regimes depending on the selection of nonlinear crystals. These results establish OPA not merely as an amplifier of structured light, but as a nonlinear platform for simultaneously scaling optical energy, translating frequency and coherently transferring spatiotemporal topology.

# 2. Results

## 2.1 Mathematical description of optical parametric amplification of spatiotemporal optical vortices

STOVs represent a fundamentally new class of structured light fields, in which optical phase singularities are embedded in the coupled spatiotemporal domain rather than solely in the transverse spatial coordinates. Unlike conventional optical vortices carrying longitudinal OAM, STOVs possess transverse OAM arising from phase circulation in a plane composed of one spatial coordinate and the local temporal coordinate.

We describe the electric field of each interacting wave as $E_j(\mathbf{r},t) = A_j(\mathbf{r},t)\exp[i(k_j z - \omega_j t)] + c.c.$, where $A_j$ denotes the slowly varying field envelope and $j = p,s,i$ represents the pump, signal and idler fields, respectively. We consider an STOV propagating along the $z$-direction, with the vortex structure defined in the $x{-}\tau$ plane, where $\tau = t - v_g z$ is the retarded time coordinate in the moving frame of the pulse. The spatiotemporal topology is encoded in the complex envelope $A_s(x,\tau)$. The complex electric field envelope of an ideal STOV can be expressed as[5]

$$A_s(x,\tau) = A_0\left(\sqrt{2}R\right)^{\ell} e^{-R^2} e^{i\ell\phi_{ST}}, \tag{1}$$

where $R = \sqrt{\left(\tau/\tau_0\right)^2 + \left(x/w_x\right)^2}$ and $\phi_{ST} = \tan^{-1}\left(\dfrac{\tau/\tau_0}{x/w_x}\right)$ denote the normalized radial and azimuthal coordinates in the spatiotemporal plane, respectively. The integer $\ell$ represents the spatiotemporal topological charge, determined by the phase circulation $\oint_C \nabla\phi_{ST} \cdot d\mathbf{l} = 2\pi\ell$. Therefore, the topological charge is a quantized property of the complex field and remains invariant under continuous transformations as long as the phase singularity is preserved.

Optical parametric amplification provides a coherent nonlinear mechanism for manipulating such topological fields through three-wave mixing in a second-order nonlinear medium. The nonlinear polarization driving the interaction is given by

$$P^{(2)} = \varepsilon_0 \chi^{(2)} E_p E_s^* + c.c., \tag{2}$$

where $E_p$, and $E_s$ represent the slowly varying electric-field envelopes of the pump, and signal fields, respectively; $\varepsilon_0$ is the vacuum permittivity, and $\chi^{(2)}$ denotes the second-order nonlinear susceptibility tensor of the crystal. Under the slowly varying envelope approximation, the coupled-wave equations describing the parametric interaction are

$$\frac{dA_s}{dz} = \kappa_s A_i^* e^{i\Delta kz}, \tag{3.1}$$

$$\frac{dA_i}{dz} = \kappa_i A_s^* e^{i\Delta kz}, \tag{3.2}$$

where $A_s$, $A_i$ and $A_p$ represent the slowly varying envelopes of the signal, idler and pump fields, respectively. The nonlinear coupling coefficients are given by $\kappa_s = \dfrac{2i\omega_s^2 d_{\mathrm{eff}}}{k_s c^2} A_p$, $\kappa_i = \dfrac{2i\omega_i^2 d_{\mathrm{eff}}}{k_i c^2} A_p$, where $d_{\mathrm{eff}}$ is the effective nonlinear coefficient of the crystal, $k_s$ and $k_i$ are the wavevectors of the signal and idler fields. The phase mismatch term is defined as $\Delta k = k_p - k_s - k_i$, which determines the efficiency and bandwidth of the parametric interaction. Under the undepleted pump approximation, no idler field is injected [ $A_i(0)=0$ ] and perfect phase matching ( $\Delta k=0$ ), the coupled equations have the analytical solutions (see Supplementary Note 1and 2)

$$A_s^{out}\left(x,\tau\right)=A_s^{in}\left(x,\tau\right)\cosh(gL), \tag{4.1}$$

$$A_i^{out}\left(x,\tau\right)=\frac{\kappa_i}{g}A_s^{in*}\left(x,\tau\right)\sinh(gL), \tag{4.2}$$

where $L$ is the nonlinear crystal length and $g=\sqrt{\kappa_s\kappa_i^*}$ is the parametric gain coefficient.

For a uniform pump envelope, the parametric gain acts as a real positive scalar amplification factor on the spatiotemporal envelope. Consequently, the amplified signal preserves the complex phase structure and the phase singularity of the incident STOV. In contrast, the generated idler field is coupled to the complex conjugate of the signal field, resulting in the transfer of a phase-conjugated spatiotemporal topology to the idler field. The phase relation of the idler field originates from the nonlinear coupling coefficient, which contains the phase of the pump field. Because the idler coupling coefficient is proportional to the pump envelope ( $\kappa_i\propto A_p$ ), the generated idler field inherits the pump phase through the nonlinear coupling coefficient. Writing $A_p=|A_p|e^{i\phi_p}$ and $A_s=|A_s|e^{i\phi_s}$ , Eq. (4.2) gives $A_i(L)\propto|A_p||A_s|e^{i(\phi_p-\phi_s)}$, which leads to the three wave phase relation $\phi_i=\phi_p-\phi_s$. Since the spatiotemporal topological charge is determined by the phase circulation around the singularity, the phase relation directly gives the topological-charge selection rule

$$\ell_i=\ell_p-\ell_s \tag{5}$$

For the topologically trivial Gaussian pump used in the present experiment, $\ell_p=0$ , and hence $\ell_i=-\ell_s$. Thus, the amplified signal preserves the topological charge of the incident STOV, whereas the generated idler acquires its phase-conjugated spatiotemporal topology. The general topological-charge transfer relation, including the case of a structured pump carrying its own spatiotemporal topology, is further analyzed in Supplementary Note 2.

## 2.2 Experimental platform for topology-preserving OPA of STOVs

To explore the nonlinear evolution of spatiotemporal optical topology, we developed a two-stage OPA platform that integrates structured-field generation, nonlinear amplification and complex-field characterization (Fig. 2). The system is driven by a femtosecond Yb-based laser, providing synchronized pump and coherent seed fields for the parametric interaction.

A broadband coherent seed is generated through nonlinear spectral broadening, from which a spectral band centered at 780 nm with a bandwidth of 10 nm is selected using a bandpass filter before being amplified in the first OPA stage. Besides increasing the signal energy for structured-field manipulation, this stage simultaneously provides a phase-correlated idler reference, which is essential for retrieving the complex field of the idler generated through nonlinear conversion in the second OPA stage. The amplified signal is then converted into STOVs using a programmable spatiotemporal pulse shaper, where controlled spatial-spectral coupling is introduced through spectral phase modulation in the $x-\omega$ domain[25]. After Fourier transformation, the shaped pulses acquire well-defined spatiotemporal phase singularities.

The generated STOVs are subsequently injected into the second OPA stage, where their nonlinear evolution is investigated. By expanding the pump relative to the STOV signal, the transverse gain distribution is made approximately uniform over the structured field, thereby suppressing spatially dependent gain distortion. The finite temporal envelope of the pump, however, still provides a temporal gain window and can reshape the amplified wave packet at sufficiently high gain (see Supplementary Note 3). In this configuration, the nonlinear crystal acts not only as an energy amplifier but also as a coherent interface for transferring spatiotemporal optical information between interacting fields.

The amplified signal and generated idler fields are characterized using spatially resolved spectral interferometry, which provides direct access to the complex electric fields[15,26,27]. The reconstructed complex spatial-spectral fields are transformed into the spatiotemporal domain, enabling simultaneous determination of intensity singularities, phase winding and topological charges. The signal field is characterized using a phase-stable reference pulse, while the idler field is reconstructed using the phase-correlated idler generated in the first OPA stage.

## 2.3 Coherent amplification and preservation of STOV topology

A central challenge for structured light is the realization of high-energy optical fields while maintaining their intrinsic spatial and temporal structures. Unlike conventional Gaussian pulses, where amplification can be described by a scalar gain factor, STOVs contain phase singularities encoded in the coupled spatiotemporal domain. Therefore, determining whether nonlinear amplification can preserve such topological information is essential for extending STOVs towards high-field applications.

Figure 3(a, b) shows the amplification of a singly charged STOV with topological charge $\ell$=1. Before amplification, the reconstructed field [Fig.3(a)] exhibits the characteristic vortex signature, including the intensity minimum at the phase singularity and the continuous $2\pi$ phase circulation around the vortex core. After the second OPA, the signal energy is increased by approximately 63-fold, while the reconstructed spatiotemporal intensity and phase distributions remain highly consistent with the input STOV. In particular, the phase singularity remains localized at the vortex center, and the phase winding direction is preserved after amplification [Fig.3(b)]. The agreement between the reconstructed amplitude and phase distributions demonstrates that the parametric interaction predominantly acts as a coherent gain process rather than introducing random phase distortion. This behavior is consistent with the analytical OPA solution in Eq. (4.1), where the amplification coefficient $cosh(gL)$ modifies the field amplitude while maintaining the phase information encoded in the input signal.

We further extend the amplification process to higher-order STOVs carrying a topological charge of $\ell$=2 [Fig. 3(c, d)]. Compared with singly charged vortices, higher-order STOVs contain a more complex phase structure. Similar to high-order spatial optical vortices, second-order STOVs are intrinsically sensitive to perturbations and may undergo splitting into two first-order vortices. Nevertheless, such local rearrangement of the phase singularity does not alter the global topological charge. After amplification, the $\ell$=2 STOV achieves an approximately 58-fold energy enhancement while preserving its global spatiotemporal topology and intensity distribution.

The preservation of both first- and second-order STOVs demonstrates that the nonlinear amplification process is not limited to a specific vortex order but can coherently amplify different levels of spatiotemporal topology. The experimental results confirm that the OPA process preserves the complex field information encoded in the STOV seed, enabling scalable amplification of structured optical wave packets.

Importantly, topology preservation does not require the complete spatiotemporal intensity profile to remain unchanged. For a finite pump pulse, the local parametric gain varies across the STOV, such that OPA acts simultaneously as a coherent amplifier and a spatiotemporal gain aperture. This nonuniform gain can reshape the amplitude envelope while leaving the vortex phase winding unchanged, provided that no additional phase singularities are created or annihilated. The observed persistence of the phase singularity therefore demonstrates the robustness of the STOV topology against gain-induced mode reshaping.

## 2.4 Experimental deviations and spatiotemporal mode reshaping during OPA

Small deviations between the simulated and experimentally reconstructed STOV fields mainly arise from residual aberrations, imperfect alignment and accumulated wavefront distortions in the optical transfer and interferometric reconstruction. These imperfections locally perturb the reconstructed amplitude and phase distributions, while the phase singularities and winding directions remain unchanged.

After amplification, an additional systematic reshaping is observed as a reorientation of the characteristic intensity minimum in the $x$−$\omega$ domain. This behavior originates from the finite temporal extent of the pump, which produces a temporally nonuniform parametric gain. Preferential amplification near the pump–signal temporal overlap narrows the effective temporal gain window and, through Fourier duality, leads to spectral broadening and anisotropic reshaping of the STOV. The observed change should therefore be interpreted as gain-induced spectral reorientation and shear rather than a rigid rotation of the entire spatiotemporal spectrum. Pump-power-dependent measurements further confirm that this deformation becomes more pronounced with increasing gain, while the phase singularity remains preserved (Supplementary Note 4 Figs. S1 and S2). Thus, OPA can reshape the spatiotemporal mode geometry without altering the underlying topological charge. A detailed model of gain narrowing under finite Gaussian pumping and its mapping into spectral reshaping is provided in Supplementary Note 3.

## 2.5 Generation of conjugated idler STOVs through topology transfer

Beyond coherent amplification of the input STOV signal, optical parametric amplification provides an additional pathway for transferring spatiotemporal topology to a newly generated optical frequency component. According to the phase relation established by the coupled-wave theory, the idler field inherits the phase-conjugated topology of the signal field. For a Gaussian pump with zero topological charge, the generated idler therefore carries an opposite topological charge to that of the signal.

To experimentally verify this topology transfer process, we reconstruct the complete spatiotemporal complex fields of the idler generated in the second OPA stage using spatially resolved spectral interferometry. The idler generated from the first OPA stage serves as a phase-stable reference field, enabling direct retrieval of the amplitude and phase distributions of the newly generated idler STOV.

Figure 4(a) shows the generation of an idler STOV corresponding to an input signal STOV with topological charge $\ell_s = +1$. In the $x-\omega$ domain, the generated idler exhibits the characteristic spatial-spectral structure associated with a spatiotemporal vortex. After Fourier transformation into the $x-t$ domain, the reconstructed field reveals the intensity singularity and phase winding expected for a first-order STOV. Importantly, the phase circulation direction is reversed compared with the source signal, confirming the phase-conjugated nature of the generated idler field. The measured idler therefore carries a topological charge of $\ell_i = -1$.

To further demonstrate that this behavior is not restricted to a single vortex order, we investigate the nonlinear generation of higher-order idler STOVs. As shown in Fig. 4(b), an input signal with $\ell_s = +2$ produces an idler field with the corresponding conjugated phase structure. The reconstructed phase distribution exhibits the reversed phase winding expected for a conjugated second-order topology, with the total topological charge preserved, corresponding to $\ell_i = -2$.

The preservation of the total topological charge together with the reversal of the phase winding demonstrates that the parametric interaction transfers the topological information encoded in the input STOV rather than merely reproducing its intensity distribution. The phase-conjugation relation gives the topological-charge selection rule ( $\ell_i = \ell_p - \ell_s$ ). For the topologically trivial Gaussian pump used here, ( $\ell_p = 0$ ), and therefore ( $\ell_i = -\ell_s$ ), consistent with the experimentally observed reversal of the idler topological charge.

Compared with the amplified signal, the idler exhibits weaker gain-dependent reshaping. This behavior originates from its distinct formation process: unlike the pre-existing signal field, the idler is generated only within the pump–signal temporal-overlap window and is therefore already temporally gated at low gain, making its normalized mode comparatively less sensitive to further increases in pump power (Supplementary Note 3 and Supplementary Figs. S1 and S2). These results establish OPA not only as an energy-scaling mechanism for structured light, but also as a coherent nonlinear platform for transferring and redistributing spatiotemporal optical topology between different frequency channels.

## 3. Discussion

Beyond demonstrating topology-preserving amplification of STOVs, our work establishes optical parametric amplification as a nonlinear framework for simultaneously controlling the energy, frequency and topology of spatiotemporally structured light. The approximately 60-fold pulse-energy enhancement achieved here demonstrates that substantial parametric gain can be applied to STOVs without destroying their underlying spatiotemporal topology. Starting from a low-energy structured seed, the amplified STOV reaches 108 nW average power (54 pJ per pulse at the 2-kHz repetition rate), corresponding to an approximately 63.5-fold pulse-energy gain under an 8 mW pump excitation. The relatively low pump-to-signal conversion efficiency in the present demonstration suggests considerable opportunities for further optimization of the amplification efficiency and output energy. Importantly, the present experiment is optimized primarily for faithful amplification and complex-field reconstruction rather than maximum energy extraction. The measurements further show that topology preservation and mode fidelity are distinct properties: a finite spatiotemporal gain distribution can reshape the intensity envelope and spectral geometry while leaving the phase singularity and topological charge intact.

The demonstrated two-stage architecture provides a direct route towards substantially higher STOV pulse energies through additional amplification stages or multipass geometries. OPA technology itself has already reached the 100-mJ class with sub-two-cycle pulses[24,28], demonstrating that the available energy scale of parametric amplification extends far beyond that explored here. More recently, dispersion-engineered multipass OPA achieved a gain approximately (1,500 $\times$ ) higher than its single-pass counterpart, together with a photon conversion efficiency of up to 81% and a system conversion efficiency of 52% while preserving the spatial beam quality[29]. These advances suggest that the present low-energy STOV seed is not a fundamental limitation: stronger seeding, multi-stage amplification and repeated interaction with the nonlinear medium could progressively bridge precisely shaped STOVs from the nanojoule regime towards the microjoule and ultimately millijoule regimes. For structured spatiotemporal fields, however, the relevant challenge is not energy extraction alone. Pump depletion, back-conversion, group-velocity mismatch, spatial walk-off, dispersion and nonuniform spatiotemporal gain can progressively modify the normalized field geometry. Future high-energy STOV amplifiers will therefore require simultaneous optimization of conversion efficiency and spatiotemporal mode fidelity.

The parametric interaction demonstrated here also provides a distinct perspective on angular-momentum conservation in spatiotemporally structured light. Orbital-angular-momentum transfer has been extensively established in nonlinear interactions of conventional spatial vortices, including stimulated parametric down-conversion, optical parametric oscillation and optical parametric amplification, where the angular momenta carried by the interacting waves obey corresponding selection rules[30–33]. For STOVs, however, topological charge alone does not uniquely determine the transverse OAM, because the OAM per photon additionally depends on the spatiotemporal geometry of the wave packet. Our measurements directly establish phase conjugation and topological-charge reversal of the generated idler, whereas the associated OAM conservation follows from the complete field transformation rather than from charge inversion alone. In the ideal uniform-gain limit, the amplified signal and its phase-conjugated idler counterpart retain the same normalized spatiotemporal geometry, suggesting oppositely directed transverse-OAM contributions from the generated signal and idler fields[33]. Together with the Manley–Rowe relation for pairwise signal–idler generation, this provides a qualitative picture consistent with conservation of the total transverse OAM of the parametric interaction. This picture extends the OAM selection rules established for spatial-vortex parametric amplification[30–33] into the spatiotemporal domain, where the pump energy and transverse OAM are partitioned between the newly generated signal and idler fields.

A further opportunity arises when the pump itself is spatiotemporally structured. Previous studies of spatially structured parametric interactions have shown that OAM carried by a vortex pump can be transferred and redistributed among the down-converted fields[30–32,34]. Extending this concept from purely spatial vortices to spatiotemporally structured pumps, the general phase relation $\ell_i = \ell_p - \ell_s$ provides an analogous route towards programmable topology conversion. Introducing a structured pump separates two complementary control channels: the pump phase determines the topological charge transferred to the generated idler, whereas its intensity distribution determines the local parametric gain and therefore reshapes the amplitude distributions of both output fields. Pump engineering could therefore enable idler states with suppressed, transferred or higher-order topological charges, while simultaneous control of the pump amplitude could tailor their spatiotemporal mode geometry. More generally, extending structured pumping from purely spatial vortices to programmable spatiotemporal fields could transform OPA from a topology-preserving amplifier into a nonlinear topology synthesizer, in which energy, frequency, mode geometry and optical topology are jointly controlled during the parametric interaction.

# 4. Conclusion

In conclusion, we have established optical parametric amplification of spatiotemporal optical vortices demonstrating approximately 60-fold pulse-energy scaling while preserving the spatiotemporal topology of STOVs. The conservation of spatiotemporal OAM enables the generated idler to inherit a conjugated spatiotemporal topology and thus becomes a source that generates STOVs at a different wavelength. Since the seed beam is selected from supercontinuum white light, the available wavelength spectrum combing both signal and idler spans from visible to near-infrared wavelength regimes.

High-energy STOVs enabled by OPA could open new opportunities in strong-field light–matter interactions, including high-harmonic generation into the extreme-ultraviolet regime[35], high-field harmonic processes with transverse-OAM transfer[16,17], and the study of ultrafast electronic dynamics[36]. OPA provides a critical bridge between precisely synthesized low-energy spatiotemporal fields and the high-field regime required for nonlinear and attosecond science. Together with the flexibility to access different wavelength channels by selecting spectral bands from the broadband seed and adjusting the phase-matching conditions, this platform empowers energy scaling, wavelength tuning, and spatiotemporal topology engineering.

# Methods

The experimental system was driven by a Yb-based femtosecond laser operating at 1030 nm, delivering pulses with an energy of approximately 100 μJ, a duration of ~200 fs and a repetition rate of 2 kHz. A small fraction of the fundamental pulse was attenuated to 3–4 μJ and focused into a YAG plate using an 80 mm focal-length lens to generate a broadband white-light continuum. The remaining fundamental pulse was frequency doubled in a 1-mm-thick type-I BBO crystal to generate the 515 nm pump pulse. Both OPA stages employed 1.5-mm-thick type-I BBO crystals for parametric amplification.

The amplified 780 nm signal was converted into STOVs using a programmable spatial–spectral pulse shaper consisting of a diffraction grating (1200 lines/mm), a cylindrical lens (100 mm focal length) and a phase-only spatial light modulator (SLM). The designed complex field in the $x$−$\omega$ domain was encoded onto the SLM using a phase-only hologram:

$$\mathrm{CGH}(x,\omega)=\mathrm{mod}\left\{\arg[\psi(x,\omega)]+f_x x\cdot \mathrm{asinc}[1-|\psi(x,\omega)|]+\pi \mathrm{asinc}[|\psi(x,\omega)|],2\pi\right\}, \quad (6)$$

where $\psi(x, \omega)$ represents the normalized target complex field in spatial-spectral domain and $f_x$ is the carrier spatial frequency for diffraction-order separation. After spectral recombination, the encoded spatial–frequency correlation generated STOVs with controlled topological charges. After pulse shaper, the programmed spatial–frequency correlation was converted into the desired spatiotemporal phase singularity, generating STOVs with controlled topological charges.

The complex fields of the amplified signal and generated idler were reconstructed using spatially resolved spectral interferometry. The structured field and a phase-stable reference field were spectrally dispersed and recorded on a CCD camera, with the two camera axes corresponding to the transverse coordinate x and frequency coordinate ω, respectively. The measured interferogram was described by

$$I(x,\omega)=|E_j(x,\omega)+E_r(x,\omega)e^{ik_x x}|^2, \quad (7)$$

where $E_j(x,\omega)$ represents the signal or idler field and $E_r(x,\omega)$ is the corresponding reference field. The transverse phase carrier $k_x$ enables separation of the interference term from the background intensity. A two-dimensional Fourier transform of the interferogram isolates one interference sideband, yielding the complex cross term

$$C_j(x,\omega)=E_j(x,\omega)E_r^*(x,\omega). \quad (8)$$

For each field, an additional measurement with an unstructured reference field was used to remove the reference phase and common phase distortions. The relative phase was therefore obtained from

$$\Delta\phi_j(x,\omega)=\arg\left[\frac{C_{j,\mathrm{structured}}(x,\omega)}{C_{j,\mathrm{reference}}(x,\omega)}\right]. \quad (9)$$

The complex spatial–spectral field was reconstructed as

$$E_j(x,\omega)=\sqrt{I_j(x,\omega)}\exp[i\Delta\phi_j(x,\omega)]. \quad (10)$$

where $j$=$s$, $i$ denotes the signal and idler fields, respectively. After calibration of the spectral coordinate, the corresponding spatiotemporal fields were obtained by inverse Fourier transformation,

$$E_j(x,\tau)=\mathrm{F}_{\omega}^{-1}\{E_j(x,\omega)\}. \quad (11)$$

The reconstructed complex fields provide the amplitude and phase distributions required to determine the intensity singularities, phase winding and topological charges of the amplified signal and generated idler STOVs.

## Funding

We acknowledge the support from the National Key Research and Development Program of China [Grant No. 2025YFF0515101 (Q.Z)] and the National Natural Science Foundation of China (NSFC) [Grant Nos. 12434012 (Q.Z.), 62535013 (Q.Z.), 62375099 (C.W.), 62335009 (C.W.)].

## Author Contributions

H.T. and C.W. contributed equally to this work. H.T. carried out the theoretical analysis, experiments and simulations. C.W. contributed to the project conception and experimental development. Q.Z. supervised the project. All authors discussed the results and contributed to the final manuscript.

## Competing interests

Authors declare no competing interests.

## Data availability

The data that support the findings of this study are available from the corresponding authors upon reasonable request.

## References

1. Shen, Y. *et al.* Roadmap on spatiotemporal light fields. *Journal of Optics* **25**, 093001 (2023).
2. Bliokh, K. Y. *et al.* Roadmap on structured waves. *Journal of Optics* **25**, 103001 (2023).
3. Forbes, A., de Oliveira, M. & Dennis, M. R. Structured light. *Nat. Photonics* **15**, 253–262 (2021).
4. Forbes, A. Structured light from lasers. *Laser & Photonics Reviews* **13**, 1900140 (2019).
5. Chong, A., Wan, C., Chen, J. & Zhan, Q. Generation of spatiotemporal optical vortices with controllable transverse orbital angular momentum. *Nature Photonics* **14**, 350–354 (2020).
6. Hancock, S. W., Zahedpour, S., Goffin, A. & Milchberg, H. M. Free-space propagation of spatiotemporal optical vortices. *Optica* **6**, 1547–1553 (2019).
7. Kondakci, H. E. & Abouraddy, A. F. Diffraction-free space–time light sheets. *Nature Photonics* **11**, 733–740 (2017).
8. Kondakci, H. E. & Abouraddy, A. F. Optical space-time wave packets having arbitrary group velocities in free space. *Nature Communications* **10**, 929 (2019).

9. Zdagkas, A. *et al.* Observation of toroidal pulses of light. *Nature Photonics* **16**, 523–528 (2022).
10. Wan, C., Cao, Q., Chen, J., Chong, A. & Zhan, Q. Toroidal vortices of light. *Nature Photonics* **16**, 519–522 (2022).
11. Shen, Y. *et al.* Optical skyrmions and other topological quasiparticles of light. *Nature Photonics* **18**, 15–25 (2024).
12. Teng, H. *et al.* Construction of optical spatiotemporal skyrmions. *Light: Science & Applications* **14**, 324 (2025).
13. Liu, X. *et al.* Spatiotemporal optical vortices with controllable radial and azimuthal quantum numbers. *Nature Communications* **15**, 5435 (2024).
14. Hancock, S., Zahedpour, S. & Milchberg, H. Transient-grating single-shot supercontinuum spectral interferometry (TG-SSSI). *Optics Letters* **46**, 1013–1016 (2021).
15. Cheng, J., Wu, Z., Sun, A. & Wan, C. Spatiotemporal optical vortices with invariant radius and width. *Laser & Photonics Reviews* **19**, e00318 (2025).
16. Chen, Z.-Y., Hu, R., Zhang, S. & Yuan, T. Relativistic high-order harmonic generation of spatiotemporal optical vortices. *Phys. Rev. A* **106**, 013516 (2022).
17. Zhang, L., Ji, L. & Shen, B. Intense harmonic generation driven by a relativistic spatiotemporal vortex beam. *High Power Laser Science and Engineering* **10**, e46 (2022).
18. Stelzer, E. H. *et al.* Light sheet fluorescence microscopy. *Nature Reviews Methods Primers* **1**, 73 (2021).
19. Shi, Y., Zhang, X., Arefiev, A. & Shen, B. Advances in laser-plasma interactions using intense vortex laser beams. *Sci. China Phys. Mech. Astron.* **67**, 295201 (2024).
20. Wu, Z., Cheng, J. & Wan, C. Optical data encoding through spatiotemporal orbital angular momentum spectral engineering. *ACS Photonics* **12**, 5017–5023 (2025).
21. Huang, S. *et al.* Spatiotemporal vortex strings. *Sci. Adv.* **10**, eadn6206 (2024).
22. Huang, J., Zhang, J., Zhu, T. & Ruan, Z. Spatiotemporal differentiators generating optical vortices with transverse orbital angular momentum and detecting sharp change of pulse envelope. *Laser & Photonics Reviews* **16**, 2100357 (2022).
23. Hauri, C. P., Schlup, P., Arisholm, G., Biegert, J. & Keller, U. Phase-preserving chirped-pulse optical parametric amplification to 17.3 fs directly from a ti: Sapphire oscillator. *Optics Letters* **29**, 1369–1371 (2004).
24. Xu, L. & Takahashi, E. J. Dual-chirped optical parametric amplification of high-energy single-cycle laser pulses. *Nature Photonics* **18**, 99–106 (2024).
25. Cao, Q., Zhang, N., Chong, A. & Zhan, Q. Spatiotemporal hologram. *Nature Communications* **15**, 7821 (2024).
26. Bowlan, P. *et al.* Crossed-beam spectral interferometry: A simple, high-spectral-resolution method for completely characterizing complex ultrashort pulses in real time. *Optics Express* **14**, 11892–11900 (2006).
27. Gui, G. *et al.* Single-frame characterization of ultrafast pulses with spatiotemporal orbital angular momentum. *ACS Photonics* **9**, 2802 (2022).

28. Xu, L. *et al.* 100-mJ class, sub-two-cycle, carrier-envelope phase-stable dual-chirped optical parametric amplification. *Optics Letters* **47**, 3371–3374 (2022).

29. Nägele, J. H., Steinle, T., Thannheimer, J., Flad, P. & Giessen, H. Dispersion-engineered multipass optical parametric amplification. *Nature* **647**, 74–79 (2025).

30. Caetano, D. *et al.* Conservation of orbital angular momentum in stimulated down-conversion. *Physical Review A* **66**, 041801 (2002).

31. Sharma, V., Chaitanya Kumar, S., Samanta, G. K. & Ebrahim-Zadeh, M. Orbital angular momentum exchange in a picosecond optical parametric oscillator. *Optics Letters* **43**, 3606–3609 (2018).

32. De Oliveira, A. *et al.* Beyond conservation of orbital angular momentum in stimulated parametric down-conversion. *Physical Review Applied* **16**, 044019 (2021).

33. Feng, R. *et al.* Femtosecond infrared optical vortex lasers based on optical parametric amplification. *High Power Laser Science and Engineering* **10**, e29 (2022).

34. Rodrigues, R. *et al.* Orbital angular momentum symmetry in a driven optical parametric oscillator. *Optics Letters* **43**, 2486–2489 (2018).

35. Martín-Hernández, R. *et al.* Extreme-ultraviolet spatiotemporal vortices via high harmonic generation. *Nature Photonics* **19**, 817–824 (2025).

36. Chen, Y. *et al.* Atomic photoionization by spatiotemporal optical vortex pulses. *Physical Review A* **107**, 033112 (2023).

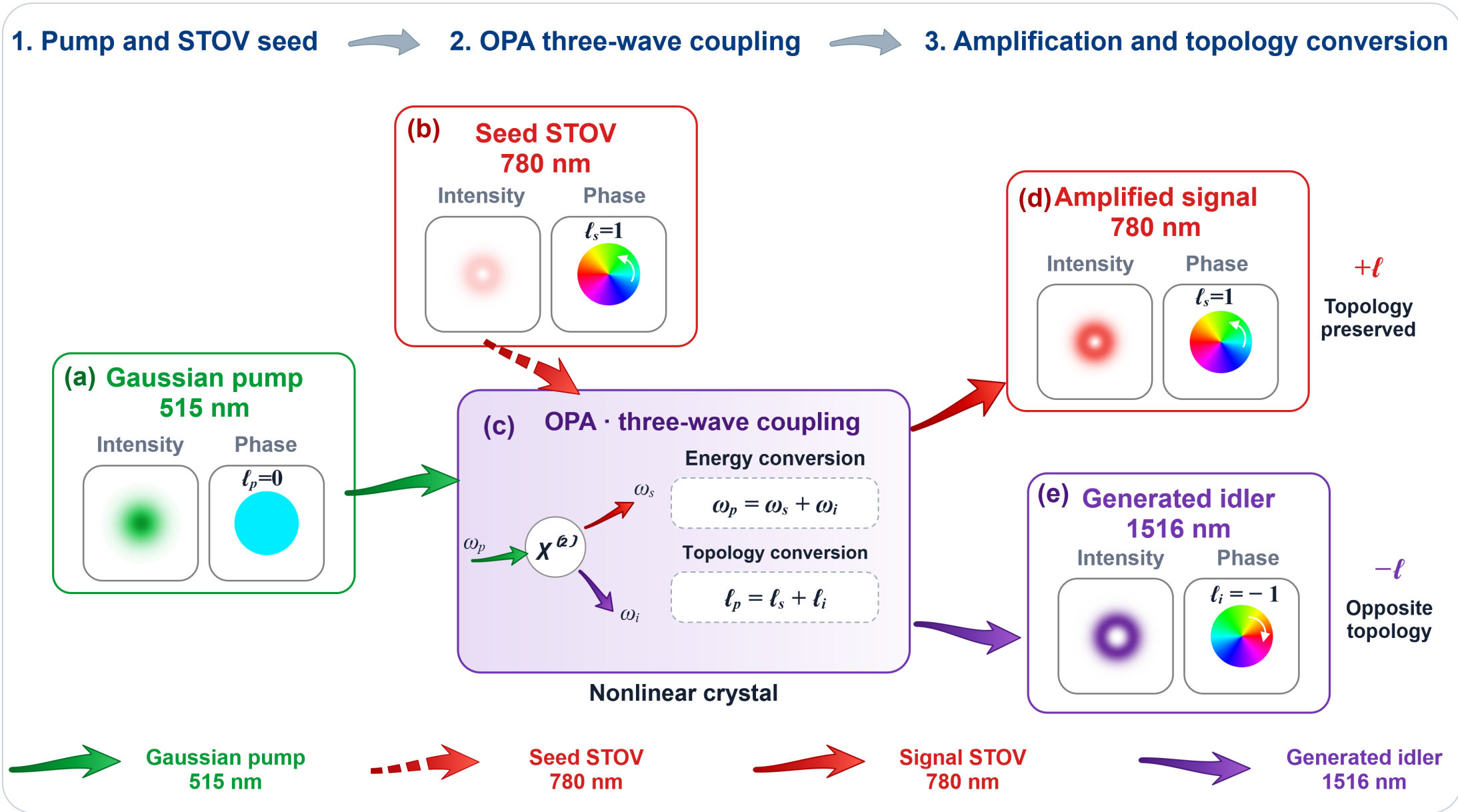

**Fig. 1 Conceptual illustration of topology-preserving optical parametric amplification of spatiotemporal optical vortices.** A spatiotemporal optical vortex interacts with a coherent pump field through a second-order nonlinear process, where optical parametric amplification enables simultaneous energy scaling and topological information transfer. **(a,b)** Complex-field representations of the Gaussian pump and the input STOV signal, respectively. The pump provides nonlinear energy transfer, while the signal carries a transverse optical topology encoded in the coupled spatiotemporal domain. **(c)** Schematic of the three-wave mixing interaction, where the pump, signal and idler fields are coherently coupled through the nonlinear medium, allowing the spatiotemporal topology to be redistributed among different frequency channels. **(d)** The amplified signal field retains the original STOV topology, demonstrating topology-preserving coherent amplification. **(e)** The generated idler field inherits the phase-conjugated topology of the input STOV. For a topologically trivial Gaussian pump, this results in an opposite topological charge of the idler field, consistent with the transverse-OAM conservation derived for the ideal parametric interaction. This concept establishes optical parametric amplification as a nonlinear platform for scaling, transferring and engineering spatiotemporal optical topology.

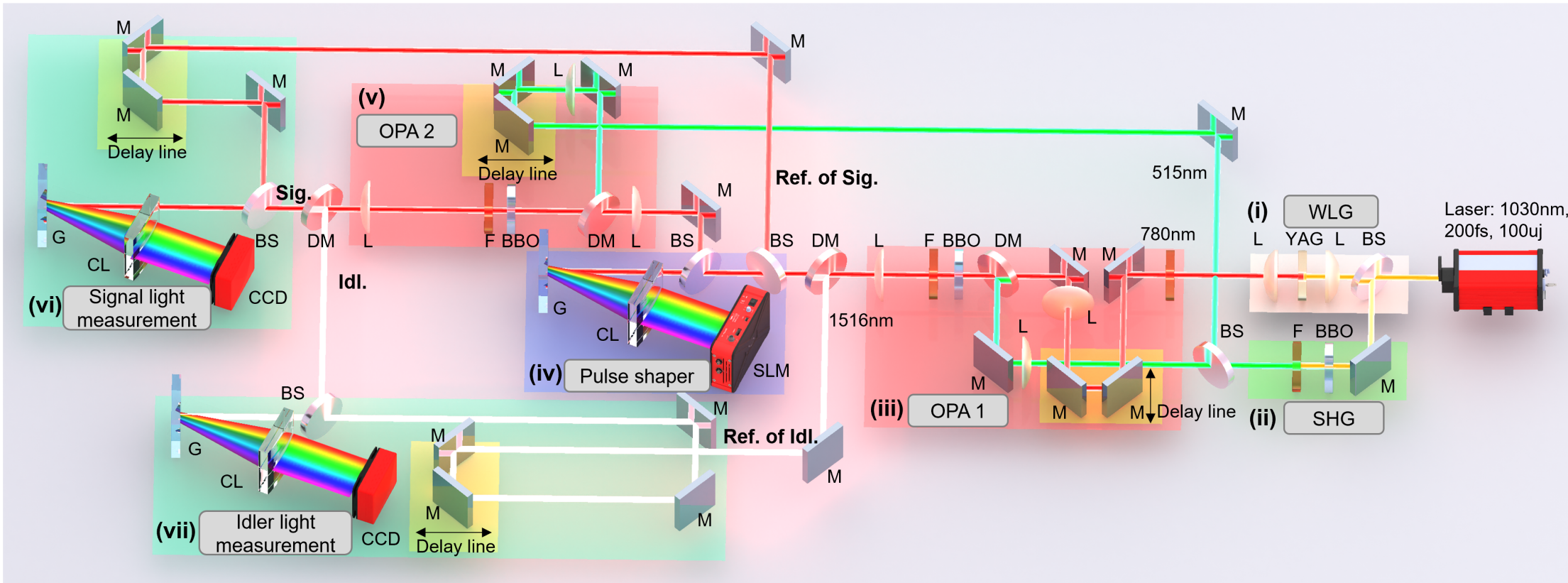

**Fig. 2 Experimental platform for generation, amplification and characterization of spatiotemporal optical vortices.** A two-stage OPA system is developed for STOV generation, amplification and complex field characterization. A broadband seed generated through supercontinuum generation **(i)** is first amplified in OPA1 **(iii)** to provide sufficient energy for spatiotemporal modulation and a coherent idler reference. The pump pulse is generated through second-harmonic generation **(ii).** A programmable pulse shaper **(iv)** consisting of a grating, cylindrical lens and spatial light modulator introduces spatial-spectral coupling to generate STOVs with controlled topological charge. The structured pulse is subsequently amplified in OPA2 **(v)** using a 515 nm pump pulse. The amplified signal and generated idler fields are fully characterized through spatially resolved spectral interferometry **(vi, vii)**, enabling reconstruction of their complex spatiotemporal electric fields. WLG, white-light generation; SHG, second-harmonic generation; OPA, optical parametric amplification; YAG, yttrium aluminum garnet; BBO, $\beta$-barium borate; M, mirror; BS, beam splitter; DM, dichroic mirror; L, lens; F, filter; SLM, spatial light modulator; G, grating; CL, cylindrical lens; CCD, charge-coupled device camera; Sig., signal; Idl., idler; Ref., reference.

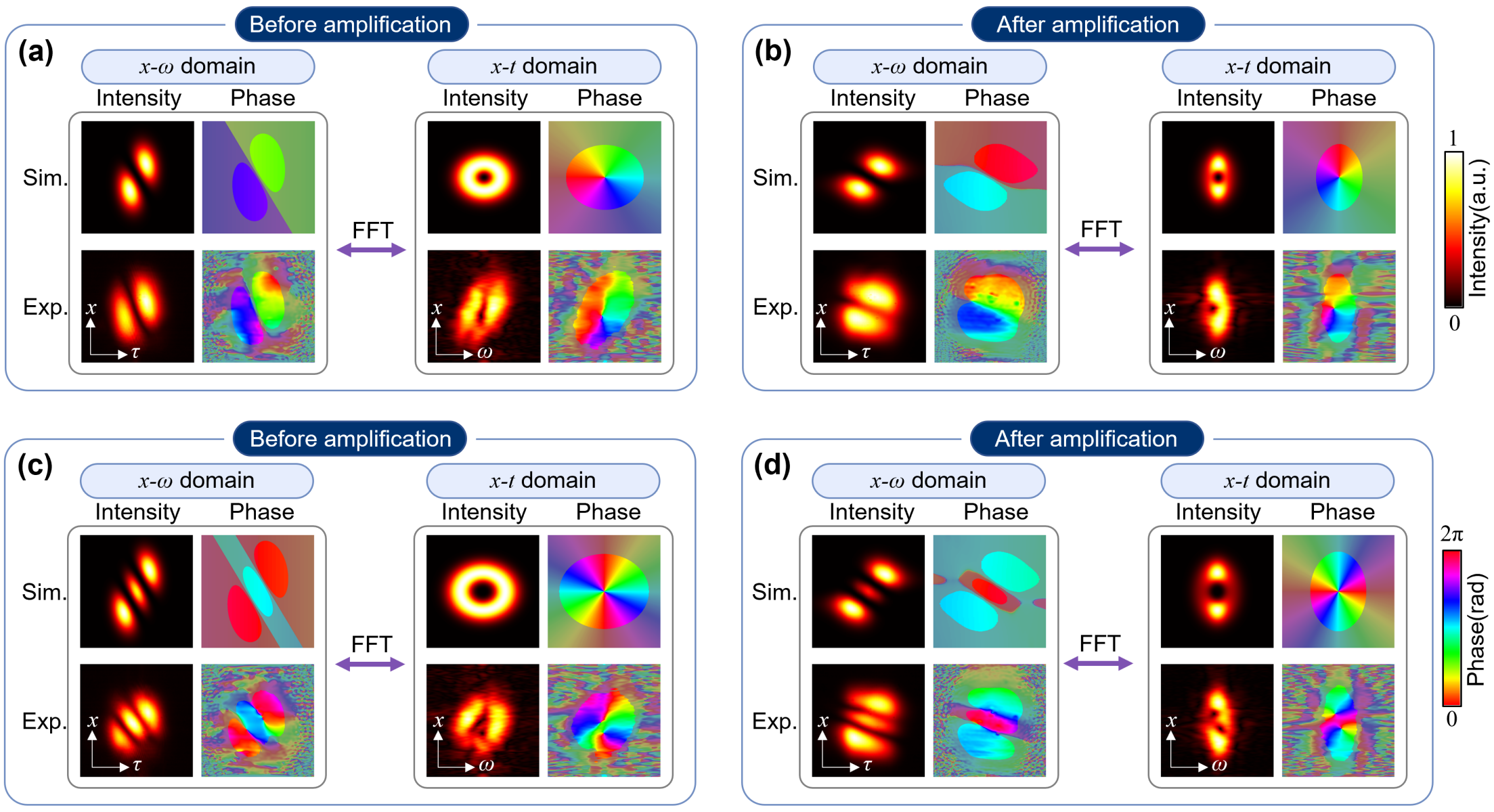

**Fig. 3 Coherent amplification and topology preservation of spatiotemporal optical vortices. (a, b)** Simulated and experimentally reconstructed complex fields of a STOV with $\ell=1$ before and after OPA. **(c, d)** Corresponding results for a STOV with $\ell=2$. For each case, intensity and phase distributions are shown in both the $x-\omega$ and $x-t$ domains, linked by Fourier transformation. The amplified fields retain the phase singularities and topological structure of the input STOVs, with measured energy gains of approximately 63× and 58× for $\ell=1$ and $\ell=2$, respectively.

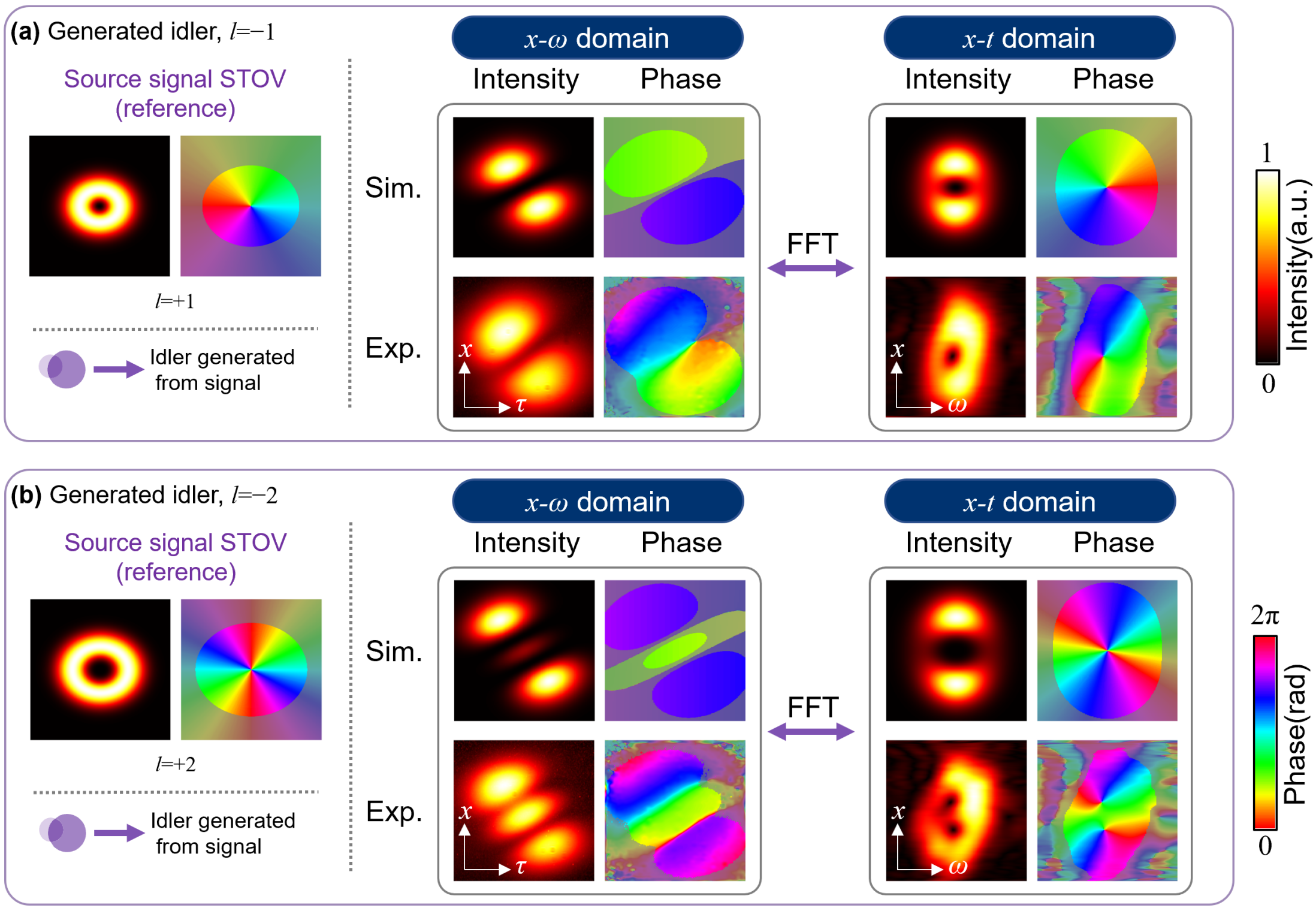

**Fig. 4 | Phase-conjugated topology transfer in generated idler STOVs. (a,b)** Simulated and experimentally reconstructed complex fields of idler STOVs generated from input signal vortices with $\ell_s = +1$ and $\ell_s = +2$, respectively. The idler fields exhibit reversed phase winding and opposite topological charges ($\ell_i = -1$ and $\ell_i = -2$) in both the $x$–$\omega$ and $x$–$t$ domains, demonstrating phase-conjugated transfer of spatiotemporal topology through nonlinear frequency conversion.